\documentclass[10pt, final, conference, twoside]{IEEEtran}
\usepackage{cite}
\usepackage{amsmath,amssymb,amsfonts}
\usepackage[margin=0.71in]{geometry}
\usepackage{array,multirow,graphicx}
\usepackage{tikz}
\usepackage{scalerel}
\usepackage{mathtools,eqparbox}
\usepackage{xcolor}
\usepackage{commath}
\usepackage{pgfplots}
\usepackage{subfigure}
\usepackage[hidelinks]{hyperref}
\usepackage[utf8]{inputenc}
\usepackage[T1]{fontenc}
\usepackage{psfrag}
\usepackage{flushend}
\def\Var{{\textrm{Var}}}
\def\E{{\textrm{E}}}

\newcommand{\lambdaris}{\lambda_{_{\mathrm{RIS}}}}
\newcommand{\C}{\cal{C}}
\newcommand{\Rg}{R_{g}}
\newcommand{\Rq}{R_{q}}

\newcommand{\lambdahaps}{\lambda_\mathrm{{HAP}}}
\newcommand{\lambdab}{\lambda_\mathrm{{B}}}

\newcommand{\HRIS}{H_{\mathrm{RIS}}}
\newcommand{\HHAP}{H_{\mathrm{HAP}}}

\def\Var{{\textrm{Var}}}
\def\E{{\mathbb{E}}}

\newtheorem{theorem}{Theorem}

\newtheorem{lemma}{Lemma}

\newtheorem{optimization problem}{Optimization Problem}
\pgfplotsset{compat=1.14}

\tikzstyle{arrow} = [thick,->,>=stealth]
\tikzstyle{block} = [rectangle, rounded corners, minimum width=1cm, minimum height=1cm,text centered, draw=black, fill=red!30]
\tikzstyle{input} = [circle, minimum width=2.5cm, minimum height=1cm, text centered, draw=black, fill=blue!30]

\usetikzlibrary{svg.path}
\definecolor{orcidlogocol}{HTML}{A6CE39}
\tikzset{
    orcidlogo/.pic={
        \fill[orcidlogocol] svg{M256,128c0,70.7-57.3,128-128,128C57.3,256,0,198.7,0,128C0,57.3,57.3,0,128,0C198.7,0,256,57.3,256,128z};
        \fill[white] svg{M86.3,186.2H70.9V79.1h15.4v48.4V186.2z}
        svg{M108.9,79.1h41.6c39.6,0,57,28.3,57,53.6c0,27.5-21.5,53.6-56.8,53.6h-41.8V79.1z M124.3,172.4h24.5c34.9,0,42.9-26.5,42.9-39.7c0-21.5-13.7-39.7-43.7-39.7h-23.7V172.4z}
        svg{M88.7,56.8c0,5.5-4.5,10.1-10.1,10.1c-5.6,0-10.1-4.6-10.1-10.1c0-5.6,4.5-10.1,10.1-10.1C84.2,46.7,88.7,51.3,88.7,56.8z};
    }
}

\newcommand\orcidicon[1]{\href{https://orcid.org/#1}{\mbox{\scalerel*{
                \begin{tikzpicture}[yscale=-1,transform shape]
                \pic{orcidlogo};
                \end{tikzpicture}
            }{|}}}}
\begin{document}

\title{Urban RIS-Assisted HAP Networks:\\Performance Analysis Using Stochastic Geometry
\vspace{-.2 cm}
}

\author{Islam~M.~Tanash\IEEEauthorrefmark{1}${\textsuperscript{\orcidicon{0000-0002-9824-6951}}}$, Ayush Kumar Dwivedi\IEEEauthorrefmark{2}${\textsuperscript{\orcidicon{0000-0003-2395-6526}}}$, and Taneli Riihonen\IEEEauthorrefmark{2}${\textsuperscript{\orcidicon{0000-0001-5416-5263}}}$\\
\IEEEauthorrefmark{1}Department of Electrical Engineering, Prince Mohammad Bin Fahd University, Saudi Arabia\\
\IEEEauthorrefmark{2}Faculty of Information Technology and Communication Sciences, Tampere University, Finland\\
\texttt{itanash@pmu.edu.sa}, \texttt{ayush.dwivedi@tuni.fi}
\vspace{-.3 cm}}

\maketitle

\begin{abstract}
This paper studies a high-altitude platform (HAP) network supported by reconfigurable intelligent surfaces (RISs). The practical irregular placement of HAPs and RISs is modeled using homogeneous Poisson point processes, while buildings that cause blockages in urban areas are modeled as a Boolean scheme of rectangles. We introduce a novel approach to characterize the statistical channel based on generalized Beta prime distribution. Analytical expressions for coverage probability and ergodic capacity in an interference-limited system are derived and validated through Monte Carlo simulations. The findings show notable performance improvements and reveal the impact of various system parameters, including blockages effect which contribute in mitigating interference from the  other visible HAPs. This proposed system could enhance connectivity and enable effective data offloading in urban environments.
\end{abstract}

\vspace{-.2 cm}
\section{Introduction}
High-altitude platform (HAP) networks have emerged as a promising solution to extend network coverage, offering long-term communication capabilities from altitudes of 20 to 50 kilometers. However, HAP communications face several challenges, including weak signal intensity due to long-distance propagation and potential blockage by buildings and obstacles. Moreover, the continuous operation of HAPs with active communication payloads leads to considerable energy consumption, which impacts their operating lifetime. This has motivated the integration of reconfigurable intelligent surfaces (RISs) with HAP networks as an efficient measure to enhance HAP network performance \cite{3D-RIS}.

RIS comprises a large array of nearly passive, programmable elements that can direct incoming signals toward specific angles, offering an energy-efficient alternative to traditional active relays. According to \cite{Ye-JPROC22:Nonterrestrial}, RISs can be integrated with HAP networks in two ways. In the first approach, an aerial RIS (ARIS) is mounted directly on the HAP. In the second approach, terrestrial RIS (TRIS) is installed on ground structures, such as rooftops or facades, to link the HAP to users. Installing an RIS on a HAP increases weight, requires more payload capacity and energy, and reduces airtime. In contrast, TRIS allows the HAP to be more agile, prolonging flight time. Moreover, TRIS infrastructure is more reusable, supporting both HAP-based and terrestrial communications.

The applications of HAPS--RIS extend beyond traditional communication, encompassing backhauling, IoT system support, and inter-HAPS links \cite{multimodeHAPS}. Recent studies have explored HAP--RIS integration by considering aspects like link budget, power and resource optimization, channel estimation, and beamforming. For instance, researchers in \cite{Alfattani-OJCOMS2021:Link} conducted a link budget analysis for RIS-enabled aerial platforms, considering both specular and scattering reflection paradigms. They derived optimal platform placement and maximum feasible reflector count, showing that HAPS-equipped RIS platforms achieve superior link performance across environments compared to other aerial and terrestrial RIS systems. The work in \cite{BeyondCell2} investigated ARIS-assisted HAP downlink communication systems, analyzing convergence properties and proposing joint power and RIS phase shift optimization schemes. Authors in \cite{BeyondCell} have focused on resource-efficient optimization strategies to maximize the number of connected users in HAP--RIS systems. Additionally, the study in \cite{AerialPlatformsRIS} has examined the performance comparison between HAPS--relay and HAPS--RIS solutions, demonstrating that HAPS--RIS offers superior energy efficiency and cost-effectiveness. Furthermore, researchers in \cite{beamformingHAPS} have investigated cooperative passive beamforming and distributed channel estimation techniques to maximize overall channel gain in RIS-aided aerial networks. The integration of multiple RIS stages has been studied in \cite{CooperativeRIS}.

Stochastic geometry is a powerful framework for modeling the dynamic nature of HAP--RIS networks. Recent works in \cite{challenge} have developed a framework that accounts for the spatial distribution of HAPs and RISs as homogeneous Poisson-process elements, offering insights into optimal system parameters. This paper leverages it, however, in a more comprehensive manner. Specifically, this paper considers an RIS-assisted HAP network with interference from other non-serving HAPS. We present a novel approach to obtain closed-form performance metrics by avoiding the complex use of the Laplace transform of interference power. We model the received signal and interference power using gamma distribution-based approximations, thereby deriving tractable closed-form expressions by approximating the signal-to-interference ratio (SIR) distribution using a generalized Beta prime distribution. Analytical expressions are derived for coverage probability and ergodic capacity for an urban environment, considering Rician fading channels. The results demonstrate the notable benefits of deploying RISs and reveal that interference from non-serving HAPs significantly degrades performance, particularly as HAP density increases. In contrast, higher RIS density enhances performance by reducing the distance to the nearest visible RIS. Our analysis also indicates that buildings can positively impact performance by effectively blocking a significant amount of interference.


\begin{figure}[!t]
    \centering
    \includegraphics[width=.9\linewidth]{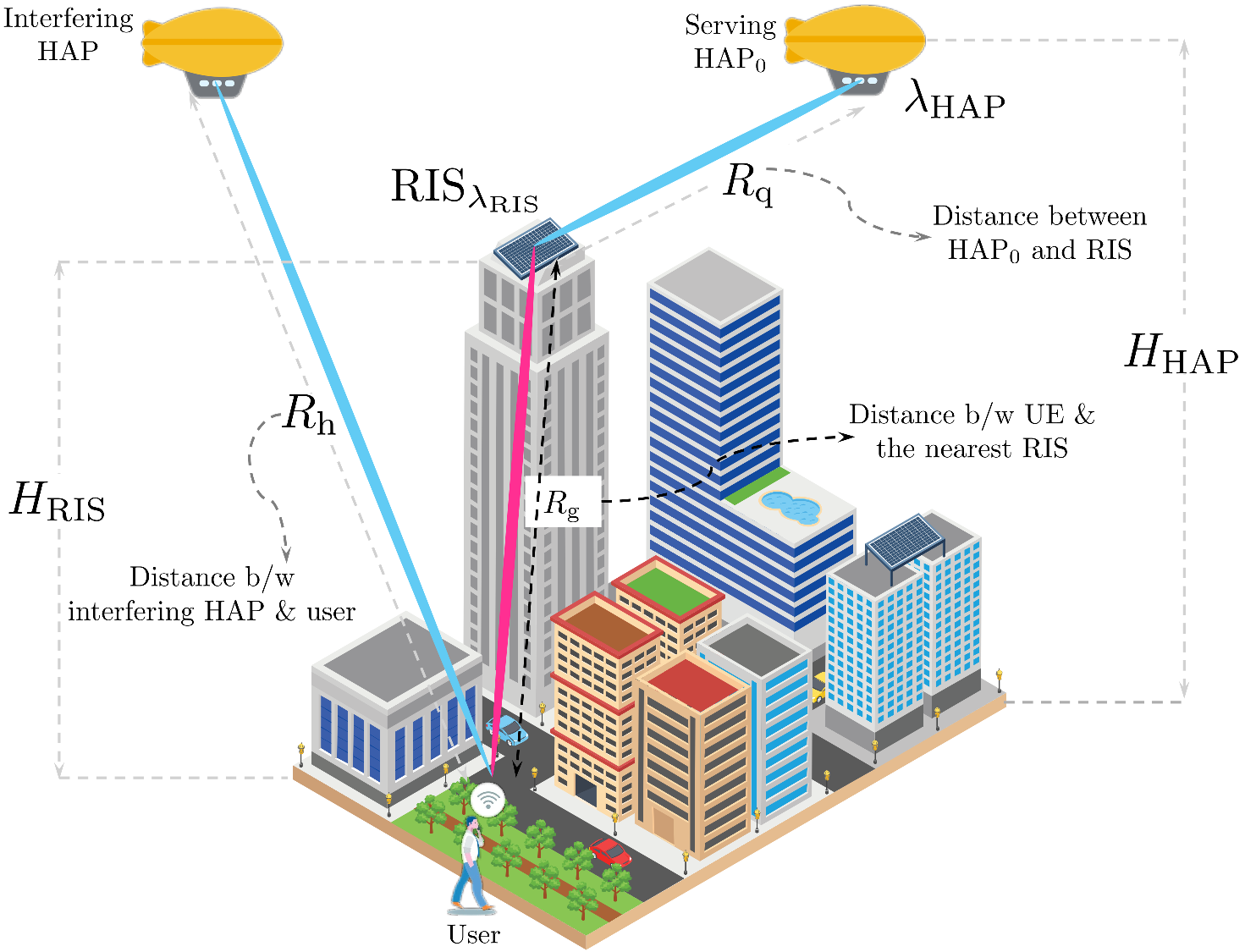}
    \caption{\textit{System model:} A network of HAPs is integrated with RISs to provide a virtual LoS communication in an urban environment. The user connects with the nearest HAP through the nearest RIS, and the RISs are deployed on buildings modeled using a Boolean model \cite{blockage_urban_main}.}
    \label{fig:systemModel}
    \vspace{-.5 cm}
\end{figure}


\section{System Model}

\subsection{RIS-Assisted HAP Network with Random Blockages}
This work considers a network of HAPs deployed in a two-dimensional (2D) space at an altitude $\HHAP$, ranging from $20$ km to $50$ km, and follows a homogeneous PPP, $\Phi$, with density $\lambdahaps$. This network is considered to operate within urban environments and is supported by RISs, which are assumed to be deployed at a height $\HRIS$, also following a homogeneous PPP, $ \Upsilon$, on an $\mathbb{R}^2$ plane with density $\lambdaris$. The RISs enhance connectivity by creating virtual line-of-sight (LoS) communication links, with each RIS equipped with $L$ reflecting elements (REs). Isotropic antennas are assumed for both HAPs and users for simplicity.

The urban environment around the user can be modeled using random shape theory, where buildings with random locations and dimensions are represented as a Boolean scheme of rectangles. Specifically, building centers are distributed across the two-dimensional plane $\mathbb{R}^2$ according to a Poisson Point Process (PPP) denoted by $\Psi$, with a density $\lambdab$. Each building $k$ is characterized by a unique set of shape parameters: length $L_k$, width $W_k$, and orientation $\Theta_k$.
The lengths $L_k$ and widths $W_k$ of the buildings are modeled as independent and identically distributed (i.i.d.) random variables. Both $L_k$ and $W_k$ are uniformly distributed, with their mean values denoted by $\mathbb{E}[L]$ and $\mathbb{E}[W]$, respectively. The orientation $\Theta_k$ of each building is a random variable uniformly distributed over the interval $(0, 2\pi]$.

\subsection{User Connectivity}

A user U connects to the nearest HAP through the closest visible RIS, as the direct HAP--U link is assumed to be blocked by buildings in the urban environment. While there may be other HAPs with a LoS to the user, these are often at greater distances and thus suffer from high path loss. In contrast, the nearest HAP, though obstructed by buildings, can provide a more efficient link through an RIS, which establishes a virtual LoS path. Additionally, connecting users only to visible HAPs could lead to uneven load distribution, with some HAPs becoming congested. By leveraging RISs to connect users to their nearest HAPs, traffic can be balanced more effectively. In a multi-HAP network, visible HAPs may still serve other applications and thus contribute to interference, making the proposed model realistic for urban scenarios. Consequently, only the HAPs visible to the user contribute to interference, rather than all HAPs in the system.

The probability that a HAP is visible to the user, as stated in \cite[Corollary 1.1]{blockage_urban_main}, is expressed as
\begin{align}
\label{eq:p_los}
\operatorname{P_{LoS}}(w_h)=\exp\left({-(\zeta w_{h}+\rho)}\right),
\end{align}   
where $\zeta=\frac{2 \lambdab(\E[L]+\E[W])}{\pi}$ and $\rho=\lambdab \, \E[L]\,\E[W]$. Accordingly, the average number of visible HAPs to a user, denoted as $M_{\text{vis}}$, is given by~\cite[Corollary 7.2]{blockage_urban_main} as
\begin{align}
M_{\text{vis}}=\frac{2\,\pi\,\lambdahaps}{\zeta^2}{\exp(-\rho)}.
\end{align}

The RISs are assumed to be positioned at elevated locations, ensuring the HAP--RIS links are unaffected by blockages. However, the RIS--U link may be blocked by buildings, so the user connects to the nearest \textit{visible} RIS instead. Throughout this paper, the nearest HAP, denoted as $\text{HAP}_o$, is referred to as the serving HAP, while the nearest visible RIS is referred to as the serving RIS.


\vspace{-.17cm}
\subsection{Signal and Fading Model}
The received signal at the ground user in an interference
limited environment can be expressed as 
\begin{align}
\label{eq:Y}
y &= \underbrace{\bigg(\frac{\sum_{l=1}^{L} \,q_{l,o}\, g_{l,o}\exp(j\theta_{l})}{\Rq^{\frac{\epsilon_q}{2}}\,\Rg^{\frac{\epsilon_g}{2}}}\bigg)s_o}_{\text{via HAP$_o$–RIS \& RIS–U link}}+ \underbrace{\hspace{-.6cm}\sum\limits_{i\in\Phi_{\textsc{LoS}}}\frac{h_i\, s_i}{R_{h,i}^{\frac{\epsilon_{h,i}}{2}}}}_{\text{via the interfering HAPs}}\hspace{-.4cm}
\end{align}
where $s_o$ is the transmitted signal of the serving HAP with power $P_0 =\mathbb{E}[|s_o|^2]$, $q_{l,o}$ and $g_{l,o}$ are the fading coefficients for HAP$_o$--RIS and RIS--U links, respectively, with $R_q$ and $R_g$ representing the corresponding distances, and $\epsilon_q$ and $\epsilon_g$ representing the corresponding path-loss exponents. The user will also experience interference from the remaining visible HAPs through the HAP$_i$–U links, each with length $R_{h,i}$. Each HAP transmits signal $s_i$ with power $P_i = \mathbb{E}[|s_i|^2]$, while the link is affected by a fading coefficient $h_i$, and a path loss exponent $\epsilon_{h,i}$.
Therefore, the interference term in (\ref{eq:Y}) accounts solely for the visible HAPs that follow nonhomogeneous PPP $\Phi_{\textsc{LoS}}$ \cite{Tanash_blockages}. Additionally, $\theta_l$ represents the adjustable phase induced by the $l$th RE. Assuming coherent signal combining and beamforming toward the intended user, interference from other HAPs via RISs is negligible. 

The maximum signal-to-interference ratio (SIR) at U can be obtained by setting $\theta_{l}=\left(\angle{q_{l,o}}+\angle{g_{l,o}}\right)$ as
\begin{align}
\label{eq:sinr}
  \text{SIR}=&\frac{P_o\left(\sum_{l=1}^{L}\left| q_{l,o}\, g_{l,o}\right|{R_q^{-\frac{\epsilon_q}{2}}\,\Rg^{-\frac{\epsilon_g}{2}}}\right)^2}{{P_i}{\sum\limits_{i\in\Phi_{\text{LoS}}}{|h_i|^2\,{R_{h,i}^{-{\epsilon_{h,i}}}}}}}=\frac{P_o\,A_{\mathcal{N}}}{P_i\,A_{\mathcal{D}}}=\frac{\mathcal{N}}{\mathcal{D}}.
\end{align}
The fading model for all links is assumed to follow a Rician distribution, as the user is expected to maintain a LoS link with the nearest visible RIS, and interference is expected only from visible HAPs. The $t$th moment of a Rician distributed fading coefficient, $\delta\in\{q_{l,o}, g_{l,o}, h_i\}$, is given as
\begin{align}
\label{eq:rician-mean}
\mathbb{E}\left[|\delta |^t\right]=\frac{\left(\sigma_{\delta}^2\right)^{\frac{t}{2}}\Gamma\left(1+\frac{t}{2}\right) \mathrm{e}^{-K_{\delta}}}{\left(K_{\delta}+1\right)^{\frac{t}{2}}} { }_1F_1\left(1+\frac{t}{2} , 1 , K_{\delta} \right).  
\end{align}
The Rician factor $K_{\delta}$ represents the ratio of the power in the LoS component to the power in the other scattered paths, whereas $\sigma_{\delta}^2 =\E\big[|\delta|^2\big]$. The above formula is acquired from \cite[Eq. 3]{kappa_mu_2} by setting $\kappa=K_{\delta}$ and $\mu=1$. 

%
\section{Statistical Characterization of the SIR}

This section derives the probability density function (PDF) of the SIR, as formulated in (\ref{eq:sinr}). The complexity of this SIR expression prevents us from obtaining a closed-form PDF, a challenge commonly encountered in the literature when interference is incorporated into the system model. This difficulty arises even in simpler scenarios involving direct links between the user and serving or interfering non-terrestrial units, without the presence of an RIS. Unlike conventional methods that derive performance measures, such as coverage probability and ergodic capacity, by substituting random parameter distributions into complex mathematical expressions, often resulting in intricate integrals and requiring the Laplace transform of interference power, which lacks a closed form, our approach bypasses these complexities. We instead derive closed-form expressions for these performance metrics by developing an accurate and tractable approximation for the SIR’s PDF, modeled with a generalized Beta prime distribution. This approach is based on approximating the received signal power and interference power with the gamma distribution first.

\subsection{Distance Distributions}
Before proceeding into the the mathematical derivation of the SIR's PDF, we present the distribution of the links' lengths, namely, HAP$_o$--RIS, RIS--U, and HAP$_i$--U links.

\subsubsection{Visible HAP$_i$--U Link}
The PDF of the horizontal distance of the visible HAPs from the user, $f_{\omega_{h,\text{LoS}}}$, is derived in the following lemma. 

\begin{lemma}
\label{lem:distance_los_hap}
The PDF of the horizontal distance corresponding to any visible HAP with a LoS path to the user is given by
\begin{align}
\label{eq:w_h_los}
 f_{\omega_{h,{\text{LoS}}}}(w_h)&=\frac{\zeta^2 w_h}{\left(1 - (\zeta  \omega_h + 1) \exp(-\zeta \omega_h)\right)}\exp\left({-\zeta w_{h}}\right), 
\end{align}
where $ \omega_h\geq\frac{\sqrt{2\exp{(-\rho)}}}{\zeta}$ is the horizontal length of the link\cite[Theorem 7]{ratio_of_gammas}.
\end{lemma}
\begin{IEEEproof}
Since any HAP can be either visible or nonvisible, we describe this event by a Bernoulli random variable denoted as $S$, where it can take values either $S=0$ or $S=1$. Therefore, a HAP is visible to the user only when $S=1$ with success probability given in (\ref{eq:p_los}), i.e., $\operatorname{Pr}(S=1\mid\omega_{h,{\text{any}}})=\operatorname{P_{LoS}}(w_h)=\exp\left({-(\zeta w_{h}+\rho)}\right)$. Accordingly, the PDF of ${\omega_{h,{\text{LoS}}}}$ which corresponds to the visible HAP, can be calculated using the joint probability formula conditioned on $0\leq w_h\leq\omega_h$, as
\begin{align}
\label{eq:proof_los_w_h}
   f_{\omega_{h,{\text{LoS}}}}(w_h)&=\frac{\operatorname{Pr}(S=1\mid \omega_{h,{\text{any}}}){f_{\omega_{h,{\text{any}}}}}(w_h)}{\operatorname{Pr}(0\leq \omega_{h,{\text{LoS}}}\leq\omega_h)},
\end{align}
with the PDF of $\omega_{h,{\text{any}}}$ given by
\begin{align}
\label{eq:w_h_any}
f_{\omega_{h,{\text{any}}}}(w_h)=2\frac{w_h}{\omega_h^2}, 0\leq \omega_h\leq\infty,  
\end{align}
which holds for any HAP--U link, regardless of visibility, and 
\begin{align}
\label{eq:cons}
&{\operatorname{Pr}(0\leq \omega_{h,{\text{LoS}}}\leq\omega_h)}=\int_0^{\omega_h}
\frac{2\,w_h}{\omega_h^2}\exp\left({-(\zeta w_{h}+\rho)}\right)\mathrm{d}w_h\nonumber\\&= \frac{2\exp(-\rho)}{\zeta^2\omega_h^2} \left(1 - (\zeta  \omega_h + 1) \exp(-\zeta \omega_h)\right).
\end{align}
Substituting (\ref{eq:p_los}), (\ref{eq:w_h_any}), and (\ref{eq:cons}) in (\ref{eq:proof_los_w_h}), yields (\ref{eq:w_h_los}).
\end{IEEEproof}

Based on (\ref{eq:w_h_los}), the $t^{\mathrm{th}}$ moment of $R_h^{-{\epsilon_h}}$, where $R_h = \sqrt{\omega_{h,{\text{LoS}}}^2 + H_{\mathrm{HAP}}^2}$, can be found as 
\begin{align}
\label{eq:mean_rh}
&\E\Big[R_h^{{-t\epsilon_h}}\Big]=  \int_{0}^{\omega_h}{ (w_h^2+\HHAP^2)^{-\frac{t\epsilon_h}{2}}} f_{\omega_{h,{\text{LoS}}}}(w_h)\,\mathrm{d}w_h\nonumber\\
& =\frac{\zeta^2}{{\left(1 - (\zeta  \omega_h + 1) \exp(-\zeta \omega_h)\right)}}\nonumber\\
&\times\int_{0}^{\omega_h}w_h{ (w_h^2+\HHAP^2)^{-\frac{t\epsilon_h}{2}}}\exp\left({-\zeta w_{h}}\right)\,\mathrm{d}w_h.
\end{align}

\subsubsection{RIS--U Link}
The user is assumed to be served by the nearest HAP with assistance from the nearest visible RIS. Due to the presence of blockages around the user, the nearest RIS may be obstructed, requiring the user to connect to the nearest visible RIS instead. 
In particular, the PDF of the horizontal distance $\omega_g$ from an arbitrarily located user $U$ to the nearest visible RIS is obtained from~\cite[Corollary 8.1]{blockage_urban_main} as
\begin{align}
\label{eq:pdf_r_gn}
f_{\omega_g}(w_g)&= 2\,\pi\,\lambdaris\,w_g\,\exp\!\left(\!-\!\left(\zeta w_g+ \rho+2\pi\lambdaris U(w_g)\right)\right),
\end{align}
where $\zeta$ and $\rho$ are defined in (\ref{eq:p_los}), and $U(w_g)$ is defined as
\begin{align}
\label{eq:U}
&U(w_g)=\frac{\exp(-\rho)}{\zeta^2}\left[1-(\zeta w_g+1)\exp(-\zeta w_g)\right].
\end{align}

Based on (\ref{eq:pdf_r_gn}), the $t^{\mathrm{th}}$ moment of $R_g^{-\frac{\epsilon_g}{2}}$, where $R_g= \sqrt{\omega_g^2 + \HRIS^2}$, is given by \cite{challenge}
\begin{equation}
\resizebox{1\hsize}{!}{$\begin{aligned}\label{eq:mean_rg}
&\E\Big[R_g^{\frac{-t\epsilon_g}{2}}\Big]= 2\,\pi\,\lambdaris \exp(-\rho)\int_{0}^{\infty}{ (w_g^2+\HRIS^2)^{-\frac{t\epsilon_g}{4}}} w_g \exp\bigg(\\
&-\Big(\zeta w_g +2\pi\lambdaris \frac{\exp(-\rho)}{(\zeta)^2}\left[1-(\zeta w_g+1)\exp(-\zeta w_g)\right]\Big)\bigg) \mathrm{d} w_g.\\
 \end{aligned}$}
\end{equation}

\subsubsection{HAP$_o$--RIS Link}

Given that RISs are much closer to the user (tens or hundreds of meters) compared to the HAP (tens or hundreds of kilometers), the horizontal distance between the HAP and RIS can be well-approximated by the distance between the nearest HAP in a homogeneous PPP and the user. This approximation holds as the RIS is assumed to be elevated, ensuring a LoS connection with the HAP, allowing us to disregard the effect of blockages. Therefore, 
the PDF of the horizontal distance $\omega_q$ is given by~\cite[Proposition 1]{challenge}:
\begin{align}\label{eq:pdf_rq}
f_{\omega_q}(w_q) &= 2\lambda_{\text{HAP}}\pi w_q \exp{(-\lambda_{\text{HAP}}\pi w_q^2)}, \,\text{for $w_q \geq 0$}.
\end{align}

The PDF in (\ref{eq:pdf_rq}) is used to derive the $t^{\mathrm{th}}$ moment of the propagation distance for the HAP$_o$--RIS link, $R_q = \sqrt{\omega_q^2 + H_{\mathrm{HAP}}^2}$, as \cite[Lemma 3]{challenge}
\begin{align}
\label{eq:mean_rq}
&\E\Big[R_q^{\frac{-t\epsilon_q}{2}}\Big]=2\lambdahaps\pi  \int_{0}^{\infty}{ (w_q^2+\HHAP^2)^{-\frac{t\epsilon_q}{4}}}\nonumber\\
& \times w_q\exp{(-\lambdahaps\pi w_q^2)} \,\mathrm{d} w_q\nonumber\\
&=\Big(\pi\,\lambdahaps\Big) ^{\frac{\epsilon_q t}{4}} e^{\pi  \HHAP^2 \lambdahaps}  \Gamma \left(1-\frac{\epsilon_q t}{4},\HHAP^2 \lambdahaps \pi \right),
\end{align}
where $\Gamma(\cdot,\cdot)$ is the upper incomplete Gamma function~\cite{tableofseries}. 

\subsection{Statistical Characterization Based on Generalized Beta Prime Distribution}

We begin by approximating both the numerator ($\mathcal{N}$) and denominator ($\mathcal{D}$) of the SIR in (\ref{eq:sinr}) with the Gamma distribution. The rationale for this approximation stems from the Central Limit Theorem (CLT), which states that the sum of independent random variables tends toward a normal distribution. Therefore, the summations in (\ref{eq:sinr}) will be normally distributed.
Consequently, $\mathcal{N}$ can be closely approximated by a non-central chi-square distribution with one degree of freedom. Hence, the distribution of $\mathcal{N}$ and $\mathcal{D}$, resembles the Gaussian PDF with a single peak, extended tail to infinity on the right, and truncated to zero on the left. Thus, both $\mathcal{N}$ and $\mathcal{D}$ can be accurately approximated by a Gamma distribution according to \cite{stochastic-book}.

In particular, $\mathcal{N}\sim \Gamma(\alpha_{\mathcal{N}}, P_o \,\beta_{\mathcal{N}})$, and $\mathcal{D}\sim \Gamma(\alpha_{\mathcal{D}}, P_i\,\beta_{\mathcal{D}})$,
with
\begin{align}
    \label{eq:alpha_beta}
    \alpha_{\mathcal{V}} =\frac{(\E[A_{\mathcal{V}}])^2}{\Var[A_{\mathcal{V}}]}
    \:\:\text{ and }\:\:
    \beta_{\mathcal{V}} =\frac{\Var[A_{\mathcal{V}}]}{\E [A_{\mathcal{V}}]},
\end{align}
for which the shorthand $\mathcal{V} \in \{\mathcal{N},\mathcal{D}\}$. The expectations for both $A_{\mathcal{N}}$ and $A_{\mathcal{D}}$ are derived respectively as 

\begin{align}
\label{eq:mean_N}
&\E[A_{\mathcal{N}}] = \E\Bigg[\Bigg(\sum_{l=1}^{L} |q_{l,o} \, g_{l,o}| \, R_q^{-\frac{\epsilon_q}{2}} \, R_g^{-\frac{\epsilon_g}{2}}\Bigg)^2\Bigg] \nonumber\\
&= \E\Bigg[\Bigg(\sum_{l=1}^{L} |q_{l,o} \, g_{l,o}|\Bigg)^2\Bigg] \, \E\big[R_q^{-\epsilon_q}\big] \, \E\big[R_g^{-\epsilon_g}\big] \nonumber\\
&= \Big(L \, \E\big[|q_{l}|^2\big] \, \E\big[|g_{l}|^2\big] + (L^2 - L) \, \E\big[|q_{l}|\big]^2 \, \E\big[|g_{l}|\big]^2\Big) \nonumber\\
&\quad \times \E\big[R_q^{-\epsilon_q}\big] \, \E\big[R_g^{-\epsilon_g}\big],
\end{align}
and
\begin{align}
\label{eq:mean_D}
&\E[A_{\mathcal{D}}] = \E\Bigg[\sum_{i \in \Phi_{\textsc{LoS}} } |h_i|^2 \, R_{h,i}^{-{\epsilon_{h,i}}}\Bigg] \nonumber\\
&= M_{\text{vis}} \, \E\big[|h_i|^2\big] \, \E\big[R_{h,i}^{-\epsilon_{h,i}}\big].
\end{align}

The variance of $A_{\mathcal{V}}$, for $\mathcal{V} \in \{\mathcal{N},\mathcal{D}\}$, is derived using the relation
\begin{align}
&\Var\Big[A_{\mathcal{V}}\Big]=\E\Big[A_{\mathcal{V}}^2\Big]-\E\Big[A_{\mathcal{V}}\Big]^2,
\end{align}
where $\E\big[A_{\mathcal{V}}\big]$ 
is derived above in (\ref{eq:mean_N}) and (\ref{eq:mean_D}), and $\E\big[A_{\mathcal{V}}^2\big]$ is derived for $\mathcal{V} \in \{\mathcal{N},\mathcal{D}\}$ as 
\begin{align}
&\E\big[A_{\mathcal{N}}^2\big] =\E\Bigg[\bigg(\sum_{l=1}^{L}\left| q_{l,o}\, g_{l,o}\right|{R_q^{-\frac{\epsilon_q}{2}}\,\Rg^{-\frac{\epsilon_g}{2}}}\bigg)^4\Bigg]\nonumber\\
&=\Bigg(L \E\big[|q_{l,o}|^4\big]\E\big[|g_{l,o}|^4\big] + 6\,\binom{L}{2} \E\big[|q_{l,o}|^2\big]^2 \E\big[|g_{l,o}|^2\big]^2\nonumber\\
& +12\,\binom{L}{1}\binom{L-1}{2} \E\big[|q_{l,o}|^2\big]\E\big[|g_{l,o}|^2\big]\E[|q_{l,o}|]^2\E[|g_{l,o}|]^2\nonumber\\
&+4\,\binom{L}{1}\binom{L-1}{1}\E\big[|q_{l,o}|^3\big]\E\big[|g_{l,o}|^3\big] E\big[|q_{l,o}|\big]\E\big[|g_{l,o}|\big]\nonumber\\
&+ 24\,\binom{L}{4}\E\big[|q_{l,o}|\big]^4\,\E\big[|g_{l,o}|\big]^4\Bigg)\,\E\big[R_q^{-{2\epsilon_q}}\big]\E\big[R_g^{-{2\epsilon_g}}\big],
\end{align}
and
\begin{align}
&\E\big[A_{\mathcal{D}}^2\big] = \E\Big[\Big({\sum\limits_{i\in\Phi_{\textsc{LoS}}}\hspace{-.25cm}{|h_i|^2\,{R_{h,i}^{-{\epsilon_{h,i}}}}}}\Big)^2\Big]\nonumber \\
&=M_{\text{vis}} \E\big[|h_i|^4\big]\E\big[R_{h,i}^{-{2\epsilon_{h,i}}}\big]\nonumber \\
& + {M_{\text{vis}} (M_{\text{vis}} - 1)} \E\big[|h_i|^2\big]^2 \E\big[R_{h,i}^{-{\epsilon_{h,i}}}\big]^2.
\end{align}
Above, $\binom{\cdot}{\cdot}$  denotes the binomial coefficient, representing combinations, and the expectations can be acquired from (\ref{eq:rician-mean}) for $\E\big[|h_i|^t\big]$, $\E\big[|g_{l,o}|^t\big]$, and $\E\big[|q_{l,o}|^t\big]$, from (\ref{eq:mean_rh}) for $\E\big[R_h^{{-t\epsilon_h}}\big]$, from (\ref{eq:mean_rg}) for $\E\big[R_g^{\frac{-t\epsilon_g}{2}}\big]$, and from (\ref{eq:mean_rq}) for $\E\big[R_q^{\frac{-t\epsilon_q}{2}}\big]$ with substituting the suitable value of $t$.  

Therefore, the SIR in (\ref{eq:sinr}), which is approximated by the ratio of two independent Gamma-distributed random variables, follows the generalized Beta prime distribution SIR$\sim\beta'(\alpha_{\mathcal{N}}, \alpha_{\mathcal{D}},1,\frac{P_o\,\beta_{\mathcal{N}}}{P_i\,\beta_{\mathcal{D}}})$ according to \cite{ratio_of_gammas}. 
Thus, the cumulative distribution function (CDF) and the PDF of the SIR are defined respectively as \cite{beta_prime_cdf} 
\begin{align}
\label{eq:cdf_sir}
&\text{F}_{\text{SIR}}\Big(x;\alpha_{\mathcal{N}} ,\alpha_{\mathcal{D}} ,1,\frac{P_o\,\beta_{\mathcal{N}}}{P_i\,\beta_{\mathcal{D}}}\Big)=\frac{\left(\frac{ {P_i\,\beta_{\mathcal{D}} \, x}}{{P_o\,\beta_{\mathcal{N}}}+ {P_i\,\beta_{\mathcal{D}} \, x}}\right)^{\alpha_{\mathcal{N}}}}{\alpha_{\mathcal{N}} B(\alpha_{\mathcal{N}}, \alpha_{\mathcal{D}})} \nonumber\\
&{ }_2F_1\left[\alpha_{\mathcal{N}}, 1-\alpha_{\mathcal{D}} ; \alpha_{\mathcal{N}}+1 ; {\frac{ {P_i\,\beta_{\mathcal{D}} \, x}}{{P_o\,\beta_{\mathcal{N}}}+ {P_i\,\beta_{\mathcal{D}} \, x}}}\right],
\end{align}
and
\begin{align}
\label{eq:pdf_sir}
&{f}_{\text{SIR}}
\Big(x;\alpha_{\mathcal{N}} ,\alpha_{\mathcal{D}} ,1,\frac{P_o\,\beta_{\mathcal{N}}}{P_i\,\beta_{\mathcal{D}}}\Big) = \frac{P_i\,\beta_{\mathcal{D}}}{P_o\,\beta_{\mathcal{N}} \, B(\alpha_{\mathcal{N}} ,\alpha_{\mathcal{D}})} \nonumber\\
&\quad \times \left(\frac{P_i\,\beta_{\mathcal{D}} \, x}{P_o\,\beta_{\mathcal{N}}}\right)^{\alpha_{\mathcal{N}} -1} \left(1 + \frac{P_i\,\beta_{\mathcal{D}} \, x}{P_o\,\beta_{\mathcal{N}}}\right)^{-(\alpha_{\mathcal{N}} + \alpha_{\mathcal{D}})},
\end{align}
where ${ }_2 F_1$ is the Gaussian hypergeometric function and $B(\cdot, \cdot)$ is the Beta function.

\section{Performance Analysis}
In this section, the coverage probability and ergodic capacity of the RIS-based HAPs' network are derived using the statistical measures derived in the previous section.

\begin{theorem}
If the user connects to the nearest HAP and is assisted by the nearest visible RIS, the coverage probability can be calculated as
\begin{align}
\label{eq:cov_prob}
\operatorname{P}_c&\simeq1-\operatorname{Pr}(\text{SIR}\leq\text{S}_{th})\nonumber\\
&=1-\Bigg(\frac{\Big(\frac{ {P_i\,\beta_{\mathcal{D}} \, \text{S}_{th}}}{{P_o\,\beta_{\mathcal{N}}}+ {P_i\,\beta_{\mathcal{D}} \, \text{S}_{th}}}\Big)^{\alpha_{\mathcal{N}}}}{\alpha_{\mathcal{N}} B(\alpha_{\mathcal{N}}, \alpha_{\mathcal{D}})}\times \nonumber\\
&{ }_2F_1\left[\alpha_{\mathcal{N}}, 1-\alpha_{\mathcal{D}} ; \alpha_{\mathcal{N}}+1 ; {\frac{ {P_i\,\beta_{\mathcal{D}} \, \text{S}_{th}}}{{P_o\,\beta_{\mathcal{N}}}+ {P_i\,\beta_{\mathcal{D}} \, \text{S}_{th}}}}\right]\Bigg),
\end{align}
where $\text{S}_{th}$ is a predefined
threshold value. The parameters $\alpha_{\mathcal{N}}$, $\alpha_{\mathcal{D}}$, $\beta_{\mathcal{N}}$, and $\beta_{\mathcal{D}}$ are defined in (\ref{eq:alpha_beta}).
\end{theorem}

\begin{theorem}
The ergodic capacity of a HAPs network assisted with RISs can be calculated as~\cite{TCOM2}
\begin{align}
\label{eq:capacity}
&\bar{\C} \simeq
\frac{\pi \, \csc(\pi \alpha_{\mathcal{D}})}{\log(2)} \Bigg( \Gamma(\alpha_{\mathcal{D}}) \left( \frac{\beta_{\mathcal{D}} P_i}{\beta_{\mathcal{N}} P_o} \right)^{-\alpha_{\mathcal{D}}}  \nonumber\\
& \times {}_2\tilde{F}_1 \left( \alpha_{\mathcal{D}}, \alpha_{\mathcal{D}} + \alpha_{\mathcal{N}}; \alpha_{\mathcal{D}} + 1; \frac{P_o \beta_{\mathcal{N}}}{P_i \beta_{\mathcal{D}}} \right) \nonumber\\
&- \frac{\beta_{\mathcal{N}} P_o  \Gamma(\alpha_{\mathcal{N}} + 1)  {}_3\tilde{F}_2 \Big(1, 1, \alpha_{\mathcal{N}} + 1; 2, 2 - \alpha_{\mathcal{D}}; \frac{P_o \beta_{\mathcal{N}}}{P_i \beta_{\mathcal{D}}}\Big)}{\beta_{\mathcal{D}} P_i \, \Gamma(\alpha_{\mathcal{D}} + \alpha_{\mathcal{N}})} \Bigg),
\end{align}
where ${}_2\tilde{F}_1$  and ${}_3\tilde{F}_2$ represent the regularized Hypergeometric function.

\begin{IEEEproof}
The ergodic capacity [bit/s/Hz] is defined as 
\begin{align}
\label{eq:cap_def}
\bar{\C} &\triangleq \E\left[\log_2(1+\text{SIR})\right]
= \int_0^\infty \log_2(1+x) \text{f}_\text{SIR}(x) \,\mathrm{d}x.
\end{align}   
Substituting the PDF derived in (\ref{eq:pdf_sir}) in (\ref{eq:cap_def}) and evaluating the integral yields (\ref{eq:capacity}).
\end{IEEEproof}
\end{theorem}

%

%
\section{Results and Observations}

This section provides a numerical validation of the generalized Beta prime distribution approximation for the SIR, using Monte Carlo simulations. We also assess system performance by analyzing the derived coverage probability and ergodic capacity, comparing these results with corresponding reference simulations. Unless otherwise specified, parameter values are set as follows: $P_o=P_i$, $\lambda_{\text{HAP}} = 5 \times 10^{-6} / \mathrm{m}^{2}$, $\mu_{\text{RIS}} = 50 \times 10^{-6} / \mathrm{m}^{2}$, $\lambdab = 100 \times 10^{-6} / \mathrm{m}^{2}$, $H_{\mathrm{HAP}} = 50~\mathrm{km}$, $H_{\mathrm{RIS}} = 50~\mathrm{m}$, $\mathbb{E}[L] = \mathbb{E}[W] = 25~\mathrm{m}$, and $\epsilon_q =\epsilon_h =\epsilon_g = 2$.

\begin{figure}[t!]
    \centering
\includegraphics[width=0.9\linewidth]{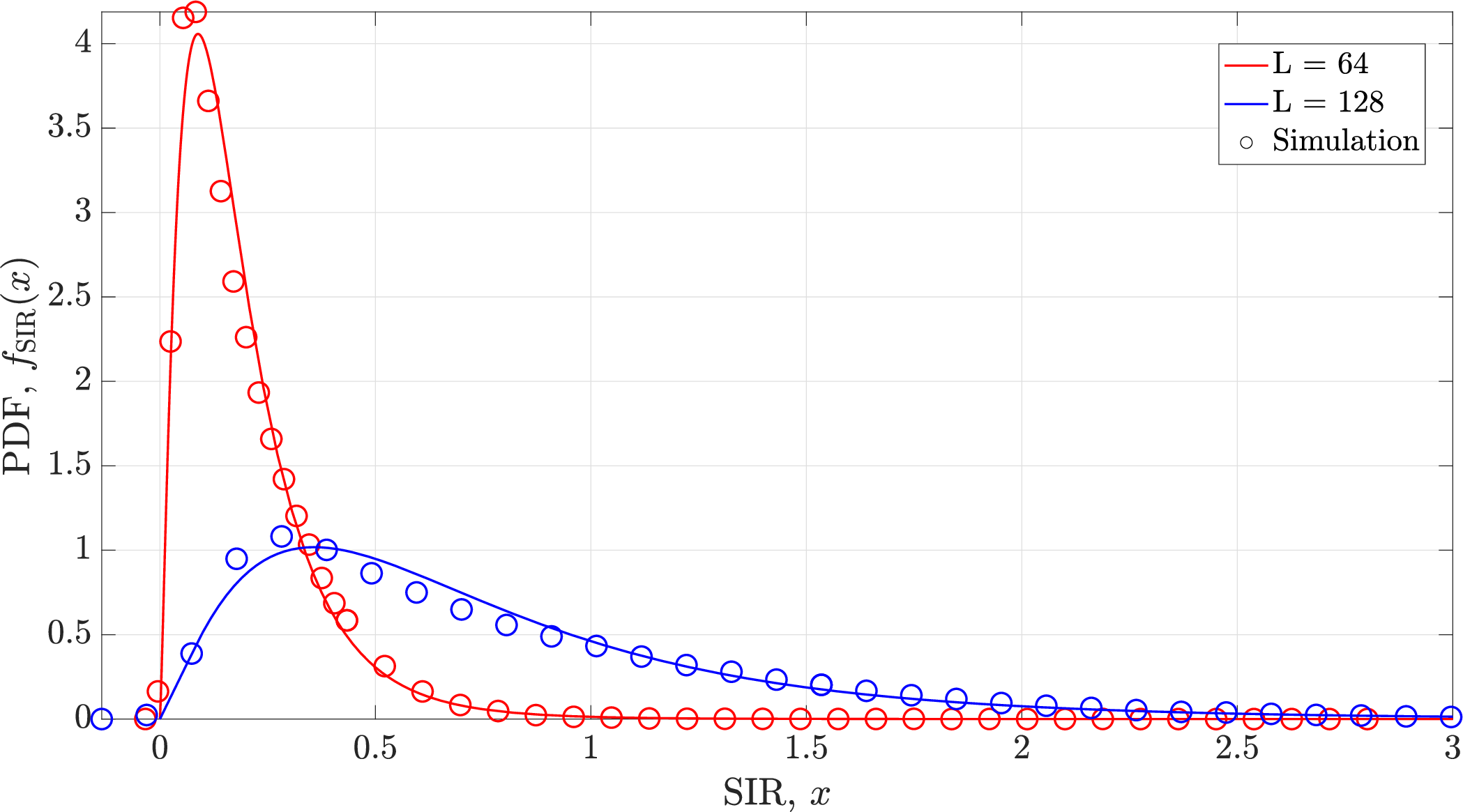}
    \vspace{-.2 cm}

    \caption{The PDF of the SIR for $L=64$ and $L=128$ REs in the RIS. The derived PDF using the generalized Beta prime distribution matches closely with the numerical PDF.}
    \label{fig:distributionPDF}
        \vspace{-.5 cm}
\end{figure}

\begin{figure}[t!]
    \centering
    \includegraphics[trim=0.8cm .2cm 1.5cm .1cm, clip=true,width=0.49\textwidth]{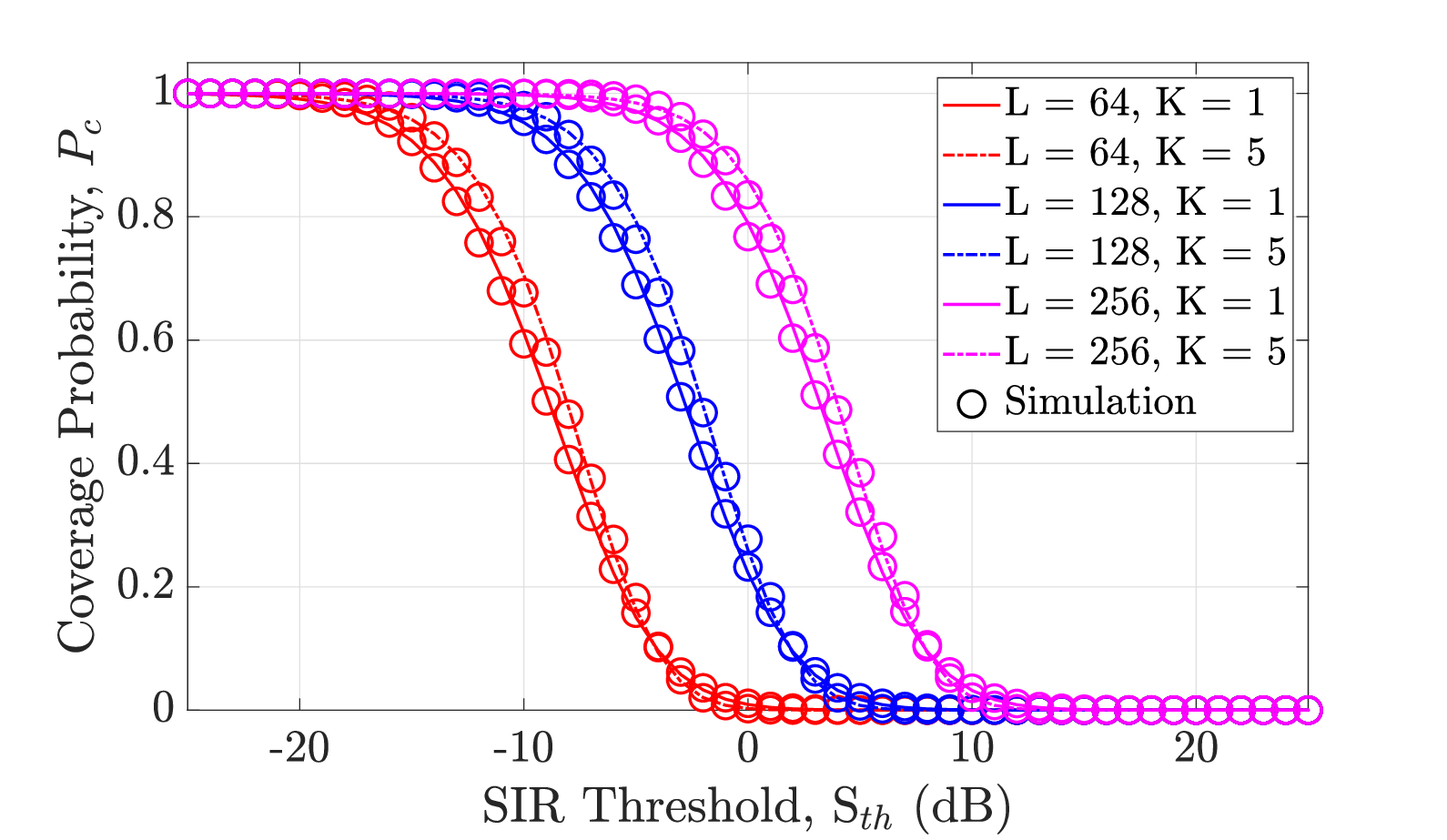}
    \caption{Coverage probability $\operatorname{P}_c$ vs. SIR threshold $\text{S}_{\text{th}}$ for varying number of reflecting elements $L$ and Rician Factor, $K$.}
    \label{fig:covSNRThreshold}
    \vspace{-.2 cm}
\end{figure}

We begin by demonstrating the high accuracy of the adopted approach in Fig. \ref{fig:distributionPDF} by plotting the simulated PDF of the SIR alongside the PDF obtained from the generalized Beta prime distribution in (\ref{eq:pdf_sir}). The results show excellent alignment between the two, for different values of REs.
This accuracy is further demonstrated in Fig. \ref{fig:covSNRThreshold}, which also illustrates the role of using RIS to enhance communication quality. A larger number of REs results in improved performance; for example, at SIR$=0$ dB and worse fading conditions ($K=1$), a coverage probability of $0.86$ is achieved with $L=256$, whereas with $L=128$, the coverage probability drops to $0.26$.

\begin{figure}[ht]
\begin{center}
{\includegraphics[trim=2.6cm .3cm .5cm .2cm, clip=true, width=.53\textwidth ]{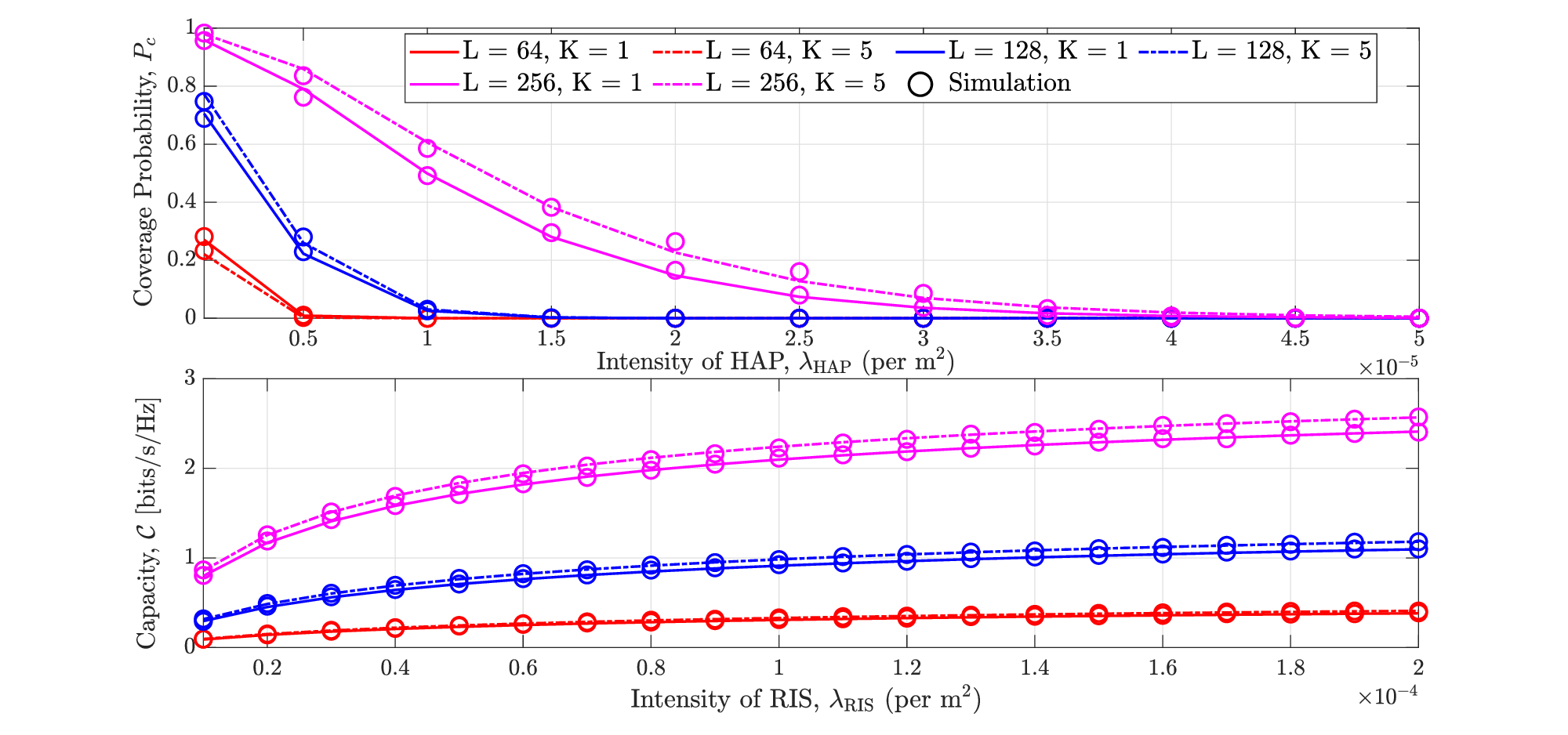}}
\caption{Coverage probability $\operatorname{P}_c$  at $\text{S}_{\text{th}}=0$ dB and ergodic capacity $\bar{\cal{C}}$ vs. intensity of RIS, $\lambdaris$ and intensity of HAP, $\lambdahaps$ for varying number of REs $L$ and Rician Factor, $K$.}
\label{fig:covCapVsLambdaRIS_LambdaHAP_T0dB}
\end{center}
    \vspace{-.4 cm}
\end{figure}

Next, the effect of the different system's parameters are studied.  Figure \ref{fig:covCapVsLambdaRIS_LambdaHAP_T0dB} shows the impact of RIS and HAP deployment density on coverage probability and ergodic capacity.  Interestingly, increasing HAP density significantly degrades coverage probability (and ergodic capacity, omitted for space) due to higher interference from the increased number of visible HAPs, which strongly affects system performance. These findings emphasize the need to account for interference in system design, highlighting the importance of this study. In addition, as RIS density increases, the distance to the nearest visible RIS decreases, enhancing ergodic capacity without introducing any interference from other visible RISs, as only the serving RIS is beamformed toward the user. However, beyond a certain density, additional RIS units no longer enhance capacity, as the environment becomes sufficiently populated for optimal coverage. A similar trend is observed in coverage probability, omitted due to space limitations.

\begin{figure}[t!]
    \centering
\includegraphics[trim=0.2cm .2cm .4cm 0cm, clip=true, width=0.49\textwidth]{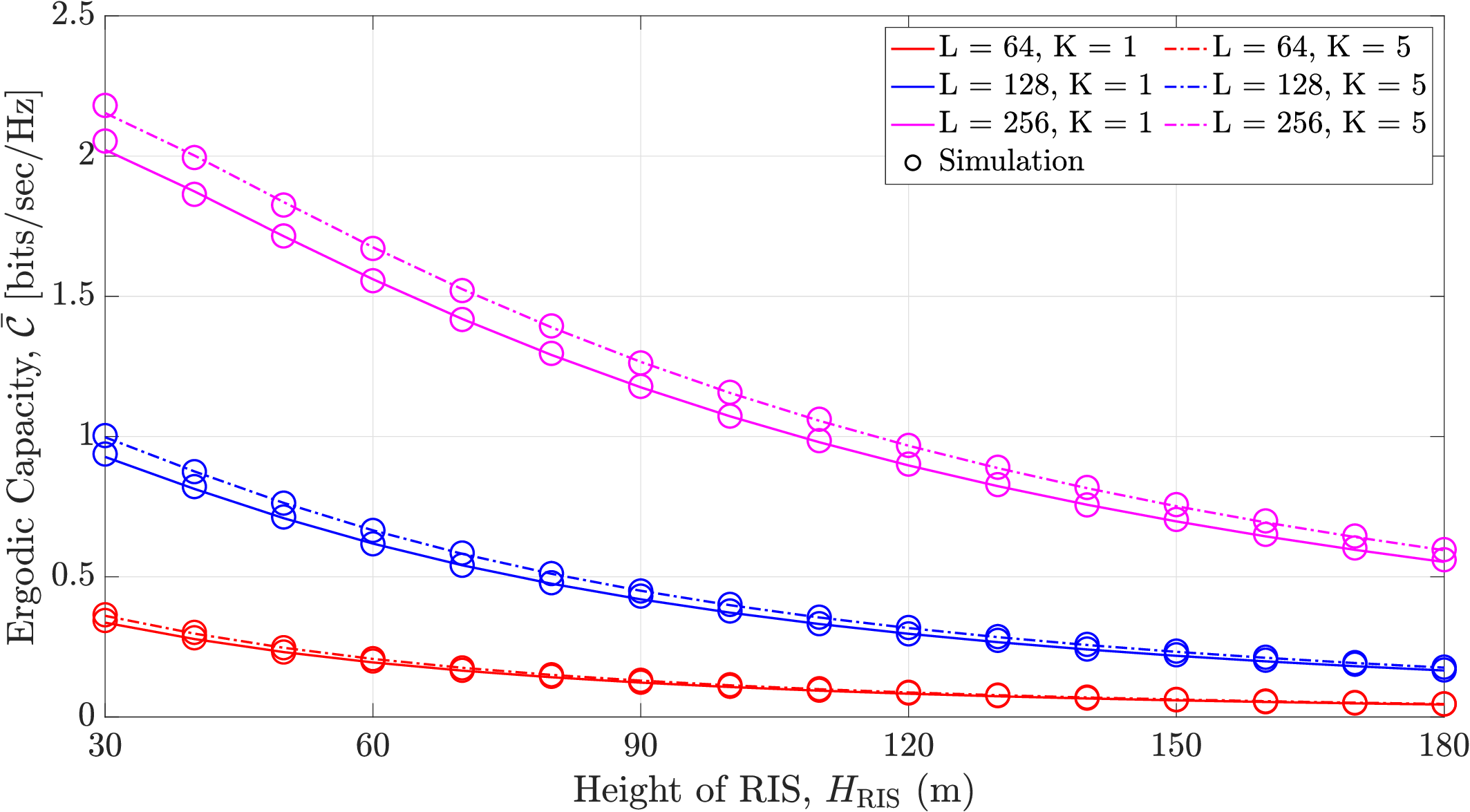}
    \caption{Ergodic capacity $\bar{\cal{C}}$ vs. Height of RIS $\HRIS$ for varying number of reflecting elements $L$ and Rician Factor, $K$.}
    \label{fig:capVshRIS}
    \vspace{-.3 cm}
\end{figure}

\begin{figure}[t!]
    \centering
\includegraphics[trim=1cm .2cm 1.5cm .05cm, clip=true, width=0.47\textwidth]{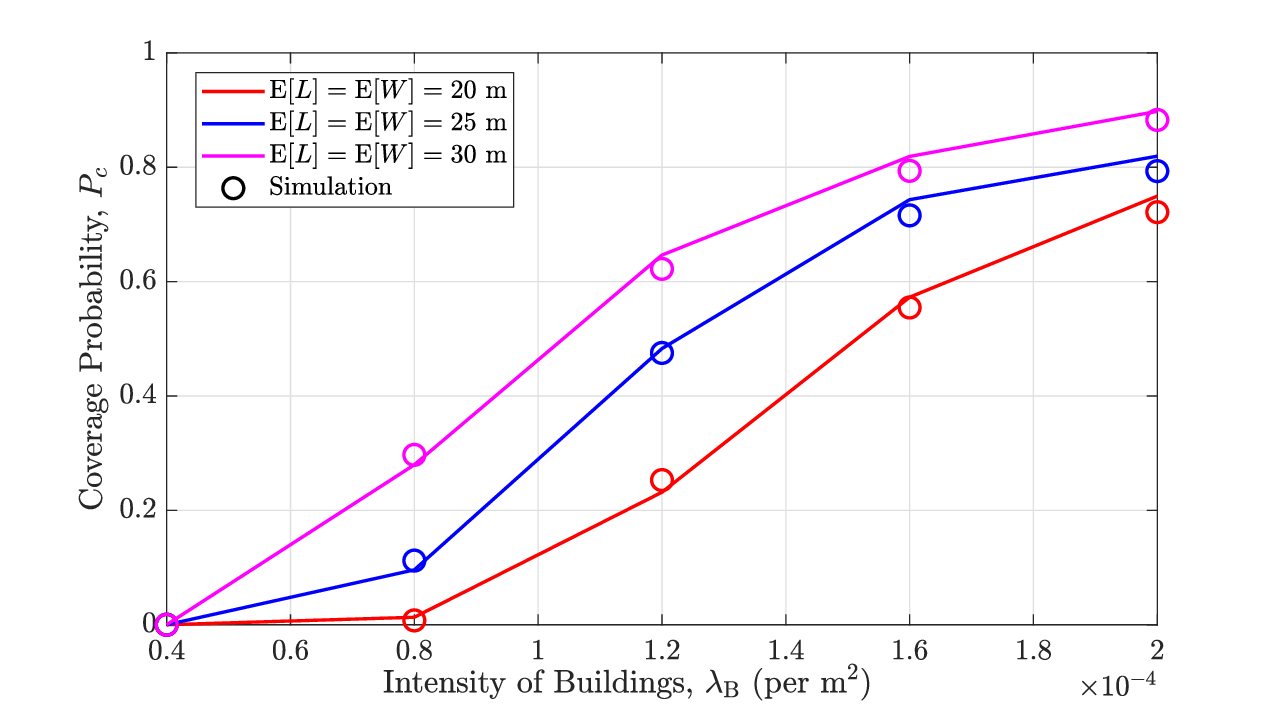}
    \caption{Coverage probability $\operatorname{P}_c$  at $\text{S}_{\text{th}}=0$ dB, $L=256$, $\lambdahaps=15 \times 10^{-6} / \mathrm{m}^{2}$, and $K=1$ vs. blockage intensity $\lambdab$ for varying building sizes.}
    \label{fig:covVslambdaB}
        \vspace{-.4 cm}
\end{figure}

In Fig. \ref{fig:capVshRIS}, we examine the impact of RIS height on performance. The results show that the ergodic capacity decreases as the RIS height increases, primarily due to the longer distances that lead to higher path losses. Therefore, placing RIS at lower altitudes is preferred to enhance performance. The impact of blockage density and size is examined in Fig. \ref{fig:covVslambdaB}. Interestingly, the figure reveals that blockages can help mitigate interference, with higher density or larger sizes further improving performance by blocking more interfering HAPs. This underscores the importance of considering blockages in the analysis and design of systems.

\section{Conclusion}
This paper studied a HAP network with RISs in urban environments, using stochastic geometry to model placements and building blockages. A novel approach with generalized Beta prime distributions was used to derive closed-form expressions for coverage probability and ergodic capacity. The study  demonstrated the significant role of RIS in enhancing system performance, especially when RIS density is optimized to reduce interference. The study also highlighted the importance of accounting for blockages effect and interference from non-serving visible HAPs in system design, as increasing HAP density can degrade performance, while increasing building density and sizes can improve performance. The impact of RIS height was also analyzed, revealing that higher placements led to increased path loss and reduced capacity. These findings emphasized the importance of interference management and optimal RIS deployment in enhancing connectivity.


\bibliographystyle{IEEEtran}
\bibliography{IEEEabrv,refs}

\begin{thebibliography}{10}
\providecommand{\url}[1]{#1}
\csname url@samestyle\endcsname
\providecommand{\newblock}{\relax}
\providecommand{\bibinfo}[2]{#2}
\providecommand{\BIBentrySTDinterwordspacing}{\spaceskip=0pt\relax}
\providecommand{\BIBentryALTinterwordstretchfactor}{4}
\providecommand{\BIBentryALTinterwordspacing}{\spaceskip=\fontdimen2\font plus
\BIBentryALTinterwordstretchfactor\fontdimen3\font minus \fontdimen4\font\relax}
\providecommand{\BIBforeignlanguage}[2]{{%
\expandafter\ifx\csname l@#1\endcsname\relax
\typeout{** WARNING: IEEEtran.bst: No hyphenation pattern has been}%
\typeout{** loaded for the language `#1'. Using the pattern for}%
\typeout{** the default language instead.}%
\else
\language=\csname l@#1\endcsname
\fi
#2}}
\providecommand{\BIBdecl}{\relax}
\BIBdecl

\bibitem{3D-RIS}
J.~He, A.~Fakhreddine, A.~S. de~Sena, Y.~Tian, and M.~Debbah, ``Unleashing {3D} connectivity in beyond {5G} networks with reconfigurable intelligent surfaces,'' in \emph{2023 57th Asilomar Conference on Signals, Systems, and Computers}, 2023, pp. 106--110.

\bibitem{Ye-JPROC22:Nonterrestrial}
J.~Ye, J.~Qiao, A.~Kammoun, and M.-S. Alouini, ``Nonterrestrial communications assisted by reconfigurable intelligent surfaces,'' \emph{Proceedings of the IEEE}, vol. 110, no.~9, pp. 1423--1465, 2022.

\bibitem{multimodeHAPS}
S.~Alfattani, W.~Jaafar, H.~Yanikomeroglu, and A.~Yongaçoglu, ``Multimode high-altitude platform stations for next-generation wireless networks: Selection mechanism, benefits, and potential challenges,'' \emph{IEEE Vehicular Technology Magazine}, vol.~18, no.~3, pp. 20--28, 2023.

\bibitem{Alfattani-OJCOMS2021:Link}
S.~Alfattani, W.~Jaafar, Y.~Hmamouche, H.~Yanikomeroglu, and A.~Yongaçoglu, ``Link budget analysis for reconfigurable smart surfaces in aerial platforms,'' \emph{IEEE Open Journal of the Communications Society}, vol.~2, pp. 1980--1995, 2021.

\bibitem{BeyondCell2}
S.~Alfattani, A.~Yadav, H.~Yanikomeroglu, and A.~Yongaçoglu, ``Beyond-cell communications via {HAPS-RIS},'' in \emph{2022 IEEE Globecom Workshops (GC Wkshps)}, 2022, pp. 1383--1388.

\bibitem{BeyondCell}
------, ``Resource-efficient {HAPS-RIS} enabled beyond-cell communications,'' \emph{IEEE Wireless Communications Letters}, vol.~12, no.~4, pp. 679--683, 2023.

\bibitem{AerialPlatformsRIS}
S.~Alfattani, W.~Jaafar, Y.~Hmamouche, H.~Yanikomeroglu, A.~Yongaçoglu, N.~D. Đào, and P.~Zhu, ``Aerial platforms with reconfigurable smart surfaces for {5G} and beyond,'' \emph{IEEE Communications Magazine}, vol.~59, no.~1, pp. 96--102, 2021.

\bibitem{beamformingHAPS}
Y.~Ni, Y.~Liu, H.~Zhao, Y.~Cai, Z.~Mo, and R.~Qiu, ``Beamforming and interference cancellation for {RIS}-assisted {HAP-D2D} communication systems,'' in \emph{2023 IEEE Globecom Workshops (GC Wkshps)}, 2023, pp. 153--159.

\bibitem{CooperativeRIS}
N.~Simmons, J.~W. Browning, S.~L. Cotton, P.~C. Sofotasios, D.~Morales-Jimenez, M.~Matthaiou, and M.~A.~B. Abbasi, ``A simulation framework for cooperative reconfigurable intelligent surface-based systems,'' \emph{IEEE Transactions on Communications}, vol.~72, no.~1, pp. 480--495, 2024.

\bibitem{challenge}
I.~M. Tanash, A.~K. Dwivedi, F.~R. Maleki, and T.~Riihonen, ``Enhancing {HAP} networks with reconfigurable intelligent surfaces,'' in \emph{Proc. 6th International Conference on Communications, Signal Processing, and their Applications (ICCSPA)}, 2024.

\bibitem{blockage_urban_main}
T.~Bai, R.~Vaze, and R.~W. Heath, ``Analysis of blockage effects on urban cellular networks,'' \emph{IEEE Transactions on Wireless Communications}, vol.~13, no.~9, pp. 5070--5083, 2014.

\bibitem{Tanash_blockages}
\BIBentryALTinterwordspacing
I.~M. Tanash and R.~Wichman, ``Performance analysis of urban satellite-terrestrial networks with {3D} blockage effects: A stochastic geometry approach,'' \emph{TechRxiv}, May 2024. [Online]. Available: \url{http://dx.doi.org/10.36227/techrxiv.171709777.76962976/v1}
\BIBentrySTDinterwordspacing

\bibitem{kappa_mu_2}
N.~Bhargav and Y.~J. Chun, ``On the product of two $\kappa$-$\mu$ random variables and its application to double and composite fading channels,'' \emph{IEEE Trans. Wireless Commun.}, vol.~17, no.~4, pp. 2457--2470, Apr. 2018.

\bibitem{ratio_of_gammas}
K.~O. Bowman, L.~R. Shenton, and P.~C. Gailey, ``Distribution of the ratio of {Gamma} variates,'' \emph{Communications in Statistics-Simulation and Computation}, vol.~27, no.~1, pp. 1--19, 1998.

\bibitem{tableofseries}
I.~Gradshteyn and I.~Ryzhik, \emph{Table of integrals, series, and products}, 7th~ed.\hskip 1em plus 0.5em minus 0.4em\relax Elsevier/Academic Press, 2007.

\bibitem{stochastic-book}
S.~Primak, \emph{\BIBforeignlanguage{eng}{Stochastic Methods and Their Applications to Communications: Stochastic Differential Equations Approach}}.\hskip 1em plus 0.5em minus 0.4em\relax Wiley, 2004.

\bibitem{beta_prime_cdf}
G.~Gallardo~i Peres, J.~Dall, P.~J. Mason, R.~Ghail, and S.~Hensley, ``A generalized beta prime distribution as the ratio probability density function for change detection between two {SAR} intensity images with different number of looks,'' \emph{IEEE Transactions on Geoscience and Remote Sensing}, vol.~62, pp. 1--14, 2024.

\bibitem{TCOM2}
I.~M. Tanash and T.~Riihonen, ``Tight logarithmic approximations and bounds for generic capacity integrals and their applications to statistical analysis of wireless systems,'' \emph{IEEE Trans. Commun.}, vol.~70, no.~10, pp. 6456--6470, 2022.

\end{thebibliography}

\end{document}